\documentstyle[12pt]{article}

\begin{document}
\begin{flushright}
hep-th/0603049\\ CAS-BHU/Preprint
\end{flushright}
\begin{center}

\vspace{0.7cm}

 {\bf \Large An alternative to the horizontality condition \\
 in superfield approach to BRST symmetries}

\vspace{1.5cm}

{\bf R. P. Malik} \footnote{On leave of absence from S. N. Bose
National Centre for Basic Sciences, Block - JD, Sector - III, Salt
Lake, Kolkata - 700 098, West Bengal, India. E-mail addresses :
malik@bhu.ac.in ; malik@bose.res.in}

{\it Centre of Advanced Studies, Physics Department,}\\ {\it
Banaras Hindu University, Varanasi - 221 005, India}

\vspace{.7cm}

\end{center}

\noindent {\bf Abstract:} We provide an alternative to the gauge
{\it covariant} horizontality condition which is responsible for
the derivation of the nilpotent (anti-)BRST symmetry
transformations for the gauge and (anti-)ghost fields of a (3 +
1)-dimensional (4D) interacting 1-form non-Abelian gauge theory in
the framework of the usual superfield approach to
Becchi-Rouet-Stora-Tyutin (BRST) formalism. The above covariant
horizontality condition is replaced by a gauge {\it invariant}
restriction on the $(4, 2)$-dimensional supermanifold,
parameterized by a set of four spacetime coordinates $x^\mu (\mu =
0, 1, 2, 3)$ and a pair of Grassmannian variables $\theta$ and
$\bar\theta$. The latter condition enables us to derive the
nilpotent (anti-)BRST symmetry transformations for {\it all} the
fields of an interacting 1-form 4D non-Abelian gauge theory where
there is an explicit coupling between the gauge field and the
Dirac fields. The key differences and striking similarities
between the above two conditions are pointed out clearly.\\



\section{Introduction}

The celebrated horizontality condition plays a key role in the
usual superfield approach [1-6] to BRST formalism when the latter
is applied to the $p$-form ($p = 1, 2, 3,...)$ (non-)Abelian gauge
theories. To be more specific and precise, in the framework of the
usual superfield approach to a given $D$-dimensional $p$-form
Abelian gauge theory, a $(p + 1)$-form super curvature $\tilde
F^{(p + 1)} = \tilde d \tilde A^{(p)}$ is constructed with the
help of the super exterior derivative $\tilde d = dx^\mu
\partial_\mu + d \theta \partial_\theta + d \bar\theta
\partial_{\bar\theta}$ (with $\tilde d^2 = 0$) and the super
$p$-form connection $\tilde A^{(p)}$ on a $(D, 2)$-dimensional
supermanifold that is parameterized by the $D$-number of commuting
spacetime variables $x^\mu$ (with $\mu = 0, 1, 2...D-1$) and a
pair of anticommuting Grassmannian variables $\theta$ and
$\bar\theta$ (i.e. $\theta^2 = \bar\theta^2 = 0, \theta \bar\theta
+ \bar\theta \theta = 0$). This super curvature is subsequently
equated to the ordinary $(p + 1)$-form curvature $F^{(p +1)} = d
A^{(p)}$ of the given $D$-dimensional Abelian $p$-form gauge
theory which is constructed with the help of the ordinary exterior
derivative $d = dx^\mu
\partial_\mu$ (with $d^2 = 0$) and the ordinary $p$-form connection
$A^{(p)}$. The process of reduction of the $(p + 1)$-form super
curvature to the ordinary $(p + 1)$-form curvature (through the
equality $\tilde F^{(p + 1)} = F^{(p + 1)}$) is known as the
horizontality condition which has been christened as the
soul-flatness condition \footnote{This condition primarily amounts
to setting equal to zero all the Grassmannian components of the
$(p + 1)$-rank (anti)symmetric curvature tensor that constitutes
the $(p + 1)$-form super curvature $\tilde F^{(p + 1)}$. The
latter is defined on the (D, 2)-dimensional supermanifold.} by
Nakanishi and Ojima [7].

The horizontality condition has also been applied to the physical
1-form non-Abelian gauge theory [3,4] where the super 2-form
curvature $\tilde F^{(2)} = \tilde d \tilde A^{(1)} + i \tilde
A^{(1)} \wedge \tilde A^{(1)}$, constructed with the help of the
super exterior derivative $\tilde d$ and the super 1-form
connection $\tilde A^{(1)}$ (by exploiting the  Maurer-Cartan
equation), is equated to the ordinary non-Abelian curvature 2-form
$F^{(2)} = d A^{(1)} + i A^{(1)} \wedge A^{(1)}$ (where the
ordinary exterior derivative $d = dx^\mu \partial_\mu$ and the
ordinary 1-form connection is $A^{(1)} = dx^\mu A_\mu$). As is
evident from our earlier discussion, the super 2-form curvature
$\tilde F^{(2)}$ is defined on the (4, 2)-dimensional
supermanifold and the ordinary 2-form curvature $F^{(2)}$ is
constructed on the ordinary 4D spacetime manifold. The key point
to be noted is that the horizontality condition is a {\it
covariant} restriction on the gauge superfield of the (4,
2)-dimensional supermanifold because the ordinary 2-form curvature
transforms covariantly under the non-Abelian gauge transformation.
This condition has been also exploited in the context of usual
superfield approach to BRST symmetries for the gravitational gauge
theories [4].

One of the most striking features of the horizontality condition
is the fact that it leads to the derivation of the nilpotent
(anti-)BRST symmetry transformations for the gauge and
(anti-)ghost fields of the Lagrangian density of an interacting
non-Abelian gauge theory. It does not shed any light, however, on
the derivation of the nilpotent (anti-)BRST symmetry
transformations associated with the matter (e.g. Dirac) fields of
the above interacting non-Abelian theory. Furthermore, it provides
the geometrical origin and interpretations for (i) the existence
of the (anti-)BRST symmetry transformations and corresponding
(anti-)BRST charges, (ii) the nilpotency property associated with
the (anti-)BRST charges (and the (anti-)BRST symmetry
transformations they generate), and (iii) the anticommutativity
property of the (anti-)BRST charges and corresponding symmetry
transformations. These beautiful geometrical interpretations,
however, remain confined to only the gauge and (anti-)ghost fields
of the (non-)Abelian theories.

The above horizontality condition has recently been augmented
[8-16] so that one could derive the nilpotent (anti-)BRST symmetry
transformations associated with  {\it all} the fields of a given
(non-)Abelian gauge and/or a reparametrization invariant theories.
These extended versions have been christened as the augmented
superfield approach to BRST formalism [8-16] where, in addition to
the horizontality condition, a set of new restrictions is imposed
on the appropriately chosen superfields of the supermanifolds. For
instance, one invokes the equality of (i) the conserved quantities
[8-13], and (ii) the gauge (i.e. BRST) invariant quantities (that
owe their origin to the (super) covariant derivatives [14-17]) in
the above extended versions of the usual superfield formalism. The
former restriction (in the case of gauge theories and
reparametrization invariant theories) leads to a logically {\it
consistent} derivation [12,13] of the nilpotent symmetry
transformations for the matter (or its analogous) fields whereas
the latter restriction, for the case of $U(1)$ and $SU(N)$ gauge
theories, yields mathematically {\it exact} nilpotent symmetry
transformations for the matter (e.g. Dirac, complex scalar) fields
[14-16]. One of the interesting features of these extensions is
the fact that the geometrical interpretations for the (anti-)BRST
symmetries and (anti-)BRST charges, found due to the application
of horizontality  condition {\it alone}, remain intact (even in
this augmented superfield formalism). However, in all the above
endeavours [8-16], one has to exploit both the restrictions (i.e.
the horizontality and the additional conditions) separately and
independently for the derivation of {\it all} the nilpotent
(anti-)BRST symmetry transformations corresponding to {\it all}
the fields of the theory.

The purpose of our present paper is to derive the on-shell as well
as off-shell nilpotent (anti-)BRST symmetry transformations for
{\it all} the fields of a specific set of Lagrangian densities of
a 4D 1-form interacting non-Abelian gauge theory by exploiting a
{\it single} gauge invariant restriction on the matter superfields
of the supermanifolds. In the process, we obtain all the results
of the horizontality condition and, on top of it, we obtain the
(anti-)BRST symmetry transformations for the matter (Dirac) fields
without spoiling the geometrical interpretations of the nilpotent
(anti-)BRST symmetry transformations (and corresponding
generators) emerging due to the horizontality condition {\it
alone}. First, as a warm up exercise, we derive the on-shell
nilpotent symmetry transformations for all the fields of a given
Lagrangian density of the 4D non-Abelian gauge theory by
exploiting a gauge invariant restriction on the chiral matter
superfields of the (4, 1)-dimensional chiral supermanifold and
pinpoint its striking similarities and key differences with the
horizontality condition. Later on, we generalize this discussion
to the general supermanifold and derive the off-shell nilpotent
(anti-)BRST transformations for all the fields of a given
non-Abelian theory. We demonstrate that the gauge (i.e. BRST)
invariant restriction on the matter superfields of the
supermanifold(s) is superior to the covariant horizontality
restriction imposed on those very supermanifold(s). To the best of
our knowledge, the BRST invariant restriction, invoked in our
present paper, has never been exploited in the context of
superfield approach to BRST formalism (except in our earlier paper
on the interacting Abelian gauge theory [17]). Thus, our present
endeavour is an important step forward in the direction of
simplifying and refining the usual superfield approach [1-7] as
well as the augmented superfield formalism [8-16] applied to the
BRST formulation of the 1-form interacting (non-)Abelian gauge
theories.

Our present paper is organized as follows. In section 2, we
discuss the bare essentials of the (anti-)BRST symmetry
transformations for the 4D 1-form interacting non-Abelian gauge
theory in the Lagrangian formulation to set up the notations and
conventions. Section 3 is devoted to the derivation of the
on-shell nilpotent BRST symmetry transformations for all the
fields of the non-Abelian theory by exploiting a gauge (i.e. BRST)
invariant restriction on the chiral matter superfields of the (4,
1)-dimensional chiral super sub-manifold. The off-shell nilpotent
(anti-)BRST symmetry transformations for all the fields are
derived in section 4 where (i) a general set of superfields are
considered on the general (4, 2)-dimensional supermanifold, and
(ii) a gauge (i.e. BRST) invariant restriction is imposed on the
matter superfields of the above supermanifold. Finally, in section
5, we make some concluding remarks, point out some key differences
between the horizontality condition and our gauge invariant
restriction and mention a few future directions for further
investigations.

\section{Preliminary: nilpotent symmetry transformations in
Lagrangian formulation}

Let us begin with the BRST invariant Lagrangian density of the
physical four $(3+1)$-dimensional non-Abelian 1-form interacting
gauge theory where there is a coupling between the gauge field and
the Dirac fields. This Lagrangian density, in the Feynman gauge,
is \footnote{ We adopt here the conventions and notations such
that the Minkowskian 4D metric $\eta_{\mu\nu} = $ diag $(+1, -1,
-1, -1)$ is flat on the spacetime manifold. The dot product and
cross product  between two non-null vectors $R^a$ and $S^a$ in the
group space of $SU(N)$ are: $R \cdot S = R^a S^a$ and $(R \times
S)^a = f^{abc} R^b S^c$, respectively. Here the Greek indices
$\mu, \nu, \rho....= 0, 1, 2, 3$ stand for the spacetime
directions on the 4D Minkowski manifold and the Latin indices $ a,
b, c.....= 1, 2, 3....$ correspond to the $SU(N)$ group indices.}
[7,18,19]
\begin{equation}
{\cal L }_b = - \frac{1}{4}F_{\mu\nu}\cdot F^{\mu\nu} +
\bar{\psi}(i\gamma^\mu D_\mu-m)\psi + B\cdot(\partial_\mu A^\mu) +
\frac{1}{2}B\cdot B - i\partial_\mu\bar{C}\cdot D^\mu C, \label{1}
\end{equation}
where $ F_{\mu\nu}= \partial_\mu A_\nu -\partial_\nu A_\mu + i
A_\mu \times A_\nu $ is the field strength tensor for the Lie
algebra valued non-Abelian gauge potential $A_\mu\equiv A_\mu ^ a
T^a$ that constitutes the 1-form $A^{(1)}$ as: $A^{(1)} = dx^\mu
A_\mu^a T^a$. Here the generators $T$'s obey the Lie algebra
$[T^a, T^b] = f^{abc} T^c$ for a given $SU(N)$ group. The
structure constant $f^{abc}$ can be chosen to be totally
antisymmetric in the indices $a, b $ and $c$ for a semi simple Lie
group $SU(N)$ [18]. The covariant derivatives $D_\mu \psi =
(\partial_\mu + i A_\mu ^a T^a) \psi $ and $D_\mu C^a =
\partial_\mu C^a + i f^{abc} A_\mu ^b C^c \equiv
\partial_\mu C^a + i (A_\mu\times C)^a$ are defined
on the matter (quark) field $\psi $ and ghost field $C^a$ such
that $[D_\mu , D_\nu]\psi= i F_{\mu\nu} \psi$ and $[D_\mu , D_\nu]
C^a = i (F_{\mu\nu} \times C)^a$. It will be noted that these
definitions for $F_{\mu\nu}$ agree with the Maurer-Cartan equation
$F^{(2)} = d A^{(1)}+ i A^{(1)}\wedge A^{(1)}\equiv \frac{1}{2!}
(dx^\mu \wedge dx^\nu) F_{\mu\nu}$ that defines the 2-form
$F^{(2)}$ which, ultimately, leads to the derivation of
$F_{\mu\nu}$. In the above equation (1), $B^a$ are the
Nakanishi-Lautrup auxiliary fields and the anticommuting (i.e.
$(C^a)^2 = (\bar{C}^a)^2 = 0, C^a \bar{C}^b + \bar{C}^b C^a =0$,
etc.) (anti-) ghost fields $(\bar C^a) C^a$ are required for the
proof of unitarity in the 1-form interacting non-Abelian gauge
theory \footnote{ For the proof of unitarity at a given order of
perturbative computation, in the context of a given physical
process involving the gauge field and the matter (quark) fields,
it turns out that for each bosonic non-Abelian gauge field (gluon)
loop diagram, a loop diagram formed by the fermionic (anti-) ghost
fields {\it alone}, is required (see, e.g. [20]).}. Furthermore,
the $\gamma$'s are the usual $4 \times 4$ Dirac matrices in the
physical 4D Minkowski space.

The above Lagrangian density (1) respects the following off-shell
nilpotent ($s_b^2 = 0$) BRST symmetry transformations ($s_b$)
[7,18,19]
\begin{eqnarray}
&&s_b A_\mu = D_\mu C,\; \qquad s_b C = - \frac{i}{2} (C \times
C),\; \qquad s_b \bar C = i B,\; \qquad s_b B = 0, \nonumber\\ &&
s_b \psi = - i (C \cdot T) \psi,\; \quad \;s_b \bar \psi = - i
\bar \psi (C \cdot T),\; \quad \;s_b F_{\mu\nu} = i (F_{\mu\nu}
\times C).\label{2}
\end{eqnarray}
The on-shell ($\partial_\mu D^\mu C = 0$) nilpotent ($\tilde s_b^2
= 0$) version of the above nilpotent symmetry transformations (
$\tilde s_b$), are
\begin{eqnarray}
&& \tilde s_b A_\mu = D_\mu C,\; \qquad \tilde s_b C = -
\frac{i}{2} (C \times C),\; \qquad \tilde s_b \bar C = - i
(\partial_\mu A^\mu), \nonumber\\ && \tilde s_b \psi = - i (C
\cdot T) \psi,\; \quad \tilde s_b \bar \psi = - i \bar \psi (C
\cdot T),\; \quad \tilde s_b F_{\mu\nu} = i (F_{\mu\nu} \times C),
\label{3}
\end{eqnarray}
under which the following Lagrangian density
\begin{eqnarray}
&& \displaystyle {{\cal L}_{b}^{(0)} = - \frac{1}{4} F^{\mu\nu}
\cdot F_{\mu\nu} + \bar \psi \;(i \gamma^\mu D_\mu - m)\; \psi}
\nonumber\\ &&- \displaystyle {\frac{1}{2} (\partial_\mu A^\mu)
\cdot (\partial_\rho A^\rho) - i
\partial_\mu \bar C \cdot D^\mu C}, \label{4}
\end{eqnarray}
changes to a total derivative (i.e. $\tilde s_b {\cal L}_b^{(0)} =
- \partial_\mu [ (\partial_\rho A^\rho) \cdot D^\mu C ])$. It is
straightforward to check that (3) and (4) are derived from (2) and
(1), respectively, by the substitution $ B = - (\partial_\mu
A^\mu)$. This relation (i.e. $B = - (\partial_\mu A^\mu)$) emerges
as the equation of motion from the Lagrangian density (1).

The following off-shell nilpotent ($s_{ab}^2 = 0$) version of the
anti-BRST ($s_{ab}$) transformations (with $s_b s_{ab} + s_{ab}
s_b = 0$):
\begin{eqnarray}
&& s_{ab} A_\mu = D_\mu \bar C,\; \qquad s_{ab} \bar C = -
\frac{i}{2} (\bar C \times \bar C),\; \qquad s_{ab} C = i \bar B,
\nonumber\\&& s_{ab} B = i (B \times \bar C),\; \qquad s_{ab}
F_{\mu\nu} = i (F_{\mu\nu} \times \bar C),  \qquad s_{ab} \bar B =
0, \nonumber\\ && s_{ab} \psi = - i (\bar C \cdot T) \psi,\;
\qquad s_{ab} \bar \psi = - i \bar \psi (\bar C \cdot T),
\label{5}
\end{eqnarray}
are the symmetry transformations for the following equivalent
Lagrangians
\begin{eqnarray}
{\cal L}_{\bar B}^{(1)} &=& - \displaystyle { \frac{1}{4}
F^{\mu\nu} \cdot F_{\mu\nu} + \bar \psi\; (i \gamma^\mu D_\mu - m
)\; \psi + B \cdot (\partial_\mu A^\mu)} \nonumber\\ &+&
\frac{1}{2} (B \cdot B + \bar B \cdot \bar B) - i \partial_\mu
\bar C \cdot D^\mu C, \label{6}
\end{eqnarray}
\begin{eqnarray}
{\cal L}_{\bar B}^{(2)} &=& - \displaystyle { \frac{1}{4}
F^{\mu\nu} \cdot F_{\mu\nu} + \bar \psi \;(i \gamma^\mu D_\mu - m
)\; \psi - \bar B \cdot (\partial_\mu A^\mu)} \nonumber\\ &+&
\frac{1}{2} (B \cdot B + \bar B \cdot \bar B) - i D_\mu \bar C
\cdot \partial^\mu C, \label{7}
\end{eqnarray}
where another auxiliary field $\bar B$ has been introduced with
the restriction $B + \bar B = - (C \times \bar C)$ (see, e.g.
[21]). It can be checked that the anticommutativity property ($s_b
s_{ab} + s_{ab} s_b = 0$) between the (anti-)BRST transformations
$s_{(a)b}$ is true for any arbitrary field of the above Lagrangian
densities. For the proof of this statement, one should also take
into account $s_b \bar B = i (\bar B \times C)$ which is not
listed in (2). We emphasize that the on-shell version of the
anti-BRST symmetry transformations, to the best of our knowledge,
does not exist for all the above cited Lagrangian densities (see,
e.g. [7,18,19]).

All types of nilpotent (of order two) symmetry transformations,
discussed and listed above, can be succinctly expressed in terms
of the conserved and off-shell nilpotent (anti-)BRST charges $Q_r$
and the on-shell nilpotent BRST charge $\tilde Q_b$, as given
below
\begin{equation}
s_r \Sigma = - i\; [\; \Sigma, Q_r\; ]_{\pm},\; \qquad\; r = b,
ab, \qquad \tilde s_b \tilde \Sigma = - i\; [\; \tilde \Sigma,
\tilde Q_b \; ]_{\pm}.\label{8}
\end{equation}
Here the $(+)-$ signs, as the subscripts on the square brackets,
stand for the brackets to be the (anti)commutator for the generic
field $\Sigma = A_\mu, C, \bar C, \psi, \bar\psi,$ $B, \bar B$ and
$\tilde \Sigma = A_\mu, C, \bar C, \psi, \bar\psi$ (present in the
above appropriate Lagrangian densities for the 1-form non-Abelian
interacting theory) being (fermionic) bosonic in nature. For our
discussions, the explicit forms of $Q_r$ (r = b, ab) and $\tilde
Q_b$  are neither essential nor urgently needed but these can be
derived by exploiting the Noether theorem (see, e.g., [7,18,19]
for details).

\section{On-shell nilpotent BRST symmetry \\transformations:
superfield approach}

In this section, first of all, we take the chiral superfields
${\cal B}_\mu^{(c)} (x, \bar\theta)$, ${\cal F}^{(c)} (x,
\bar\theta)$, $\bar {\cal F}^{(c)} (x, \bar\theta)$, $\Psi^{(c)}
(x, \bar\theta)$, $\bar \Psi^{(c)} (x, \bar\theta)$, defined on
the $(4, 1)$-dimensional super sub-manifold of the general (4,
2)-dimensional supermanifold,  as the generalization of the basic
local fields $A_\mu (x), C (x), \bar C (x), \psi (x), \bar \psi
(x)$ of the Lagrangian density (4) which are defined on the 4D
ordinary spacetime manifold. The super expansion of these chiral
superfields, in terms of the above basic local fields of the
Lagrangian density (4), are as follows
\begin{eqnarray}
({\cal B}_\mu^{(c)} \cdot T) (x, \bar\theta) &=& (A_\mu \cdot T)
(x)\; + \;\bar \theta\; (R_\mu \cdot T) (x), \nonumber\\ ({\cal
F}^{(c)} \cdot T) (x, \bar\theta) &=& (C \cdot T) (x)\; + \;i \;
\bar\theta \;(B_1 \cdot T) (x), \nonumber\\ (\bar {\cal F}^{(c)}
\cdot T) (x, \bar\theta) &=& (\bar C \cdot T) (x)\; +\; i\; \bar
\theta \;(B_2 \cdot T) (x), \nonumber\\ \Psi^{(c)} (x, \bar\theta)
&=& \psi (x) + i\;\bar \theta\; (b_1 \cdot T) (x), \nonumber\\
\bar \Psi^{(c)} (x, \bar\theta) & =& \bar \psi (x) + i\; \bar
\theta\; (b_2 \cdot T) (x). \label{9}
\end{eqnarray}
It is evident that, in the limit $\bar \theta \to 0$, we retrieve
the basic local fields of the Lagrangian density (4). In the above
expansion, there are Lie algebra valued secondary fields $R_\mu,
B_1, B_2, b_1, b_2$ which will be determined, in terms of the
basic local fields of the Lagrangian density (4), by the gauge
invariant restriction (see, e.g., equation (10) below) on the
chiral matter superfields. It will be noted that it is only the
matter fields $(\psi (x), \bar\psi (x)$ of the Lagrangian density
(4) and their chiral superfield generalizations $\Psi^{(c)} (x,
\bar\theta),$ $\bar \Psi^{(c)} (x, \bar\theta)$ that are not Lie
algebra valued. On the r.h.s. of the above expansion, all the
fields are well-behaved local fields because they are functions of
the 4D coordinates $x^\mu$ alone. Finally, the expansions in (9)
are such that the bosonic and fermionic degrees of freedom of the
local fields do match. This is an essential requirement for the
sanctity of a supersymmetric field theory.

To derive the on-shell nilpotent BRST symmetry transformations (3)
for all the local fields, present in the Lagrangian density (4),
we begin with the following gauge (i.e. BRST) invariant
restriction on the matter chiral superfields of the (4,
1)-dimensional chiral super sub-manifold:
\begin{equation}
\bar \Psi^{(c)} (x, \bar\theta)\; \tilde {\cal D}_{|(c)} \;\tilde
{\cal D}_{|(c)}\; \Psi^{(c)} (x, \bar\theta)\; = \;\bar \psi (x)\;
D \;D \; \psi (x), \label{10}
\end{equation}
where (i) the chiral super sub-manifold is parameterized by four
bosonic spacetime coordinates $x^\mu (\mu = 0, 1, 2, 3)$ and a
single Grassmannian variable $\bar\theta$, (ii) the ordinary
covariant derivative $D = dx^\mu (\partial_\mu + i A_\mu \cdot T)$
(on the r.h.s. of (10)) is defined on the ordinary 4D spacetime
manifold, (iii) the chiral super covariant derivative is: $\tilde
{\cal D}_{|(c)} = \tilde d_{|(c)} + i \tilde A^{(1)}_{|(c)}$. Here
the individual terms, present in the definition of chiral super
covariant derivative $\tilde {\cal D}_{|(c)}$, are
\begin{equation}
\tilde d_{|(c)} = dx^\mu\; \partial_\mu + d \bar\theta\;
\partial_{\bar\theta},\; \qquad \tilde A^{(1)}_{|(c)} = dx^\mu\;
{\cal B}_\mu^{(c)} (x, \bar\theta) + d \bar\theta\; {\cal F}^{(c)}
(x,\bar\theta), \label{11}
\end{equation}
(iv) the explicit computation of the r.h.s. of equation (10), on
the ordinary 4D spacetime manifold, leads to
\begin{eqnarray}
&& \bar \psi (x)\; D\; D\; \psi (x)\; = \;i \;\bar \psi (x)\;
F^{(2)}\; \psi (x), \nonumber\\ && F^{(2)} = \frac{1}{2!} \bigl
(dx^\mu \wedge dx^\nu \bigr )\; \bigl (\partial_\mu A_\nu -
\partial_\nu A_\mu + i A_\mu \times A_\nu \bigr), \label{12}
\end{eqnarray}
which is a gauge invariant quantity under the $SU(N)$ non-Abelian
transformations: $ \psi \to U \psi, \bar \psi \to \bar\psi U^{-1},
F^{(2)} \to U F^{(2)} U^{-1}$ where $U \in SU(N)$, and (v) the
definitions (11) are the chiral limit (i.e. $\bar\theta \to 0$) of
the general expressions for the super exterior derivative $\tilde
d = dx^\mu \partial_\mu + d \theta \partial_\theta + d \bar\theta
\partial_{\bar\theta}$ and super 1-form connection $ \tilde
A^{(1)} = dx^\mu {\cal B}_\mu (x, \theta, \bar \theta) + d \theta
\bar {\cal F} (x, \theta, \bar\theta) + d \bar\theta {\cal F} (x,
\theta, \bar\theta)$ defined on the general (4, 2)-dimensional
supermanifold (cf. section 4 below).

It is clear from (12) that the r.h.s. of the gauge invariant
restriction (10) yields only the coefficient of the 2-form
differential $(dx^\mu \wedge dx^\nu)$. The expansion of the l.h.s.
would, however, lead to the coefficients of all the possible
2-form differentials on the (4, 1)-dimensional chiral super
sub-manifold. The explicit form of the expansion, on the l.h.s. of
(10), yields
\begin{eqnarray}
&& (dx^\mu \wedge dx^\nu)\;\bar \Psi^{(c)}\; (\partial_\mu + i\;
{\cal B}_\mu^{(c)})\; (\partial_\nu + i\; {\cal B}^{(c)}_\nu)\;
\Psi^{(c)} + (dx^\mu \wedge d\bar\theta)\;\nonumber\\ && \bar
\Psi^{(c)} \Bigl [ (\partial_{\bar\theta} + i \;{\cal F}^{(c)})
(\partial_\mu + i\; {\cal B}_\mu^{(c)}) - (\partial_\mu + i\;
{\cal B}_\mu^{(c)}) (\partial_{\bar\theta} + i\; {\cal F}^{(c)})
\; \Bigr ]\; \Psi^{(c)} \nonumber\\ && - (d \bar\theta \wedge d
\bar\theta)\; \bar \Psi^{(c)}\; (\partial_{\bar\theta} + i\; {\cal
F}^{(c)}) (\partial_{\bar\theta} + i\; {\cal F}^{(c)})\;
\Psi^{(c)}. \label{13}
\end{eqnarray}
For algebraic convenience, it is advantageous to first focus on
the explicit computation of the coefficient of $(d \bar\theta
\wedge d \bar\theta)$. This is
\begin{equation}
- (d \bar\theta \wedge d \bar\theta)\; \bar \Psi^{(c)}\; \Bigl [\;
i \partial_{\bar\theta} {\cal F}^{(c)} - {\cal F}^{(c)} {\cal
F}^{(c)}\; \Bigr ]\; \Psi^{(c)}. \label{14}
\end{equation}
It is clear from the restriction (10) that the above coefficient
should be set equal to zero. For $\Psi^{(c)} (x, \bar\theta) \neq
0, \bar \Psi^{(c)} (x, \bar\theta) \neq 0$, we have the following
\begin{equation}
\partial_{\bar\theta} {\cal F}^{(c)} + \frac{i}{2} \; \{ {\cal
F}^{(c)}, {\cal F}^{(c)} \} = 0. \label{15}
\end{equation}
Substituting the values from the chiral expansion (9) into the
above expression, we obtain the following
\begin{equation}
i B_1 + \frac{i}{2} (C \times C) + \bar\theta (C \times B_1) = 0
\Rightarrow B_1 = - \frac{1}{2} (C \times C),\; \qquad (B_1 \times
C) = 0. \label{16}
\end{equation}
It is straightforward to note that, not only the condition $(B_1
\times C) = 0$ is satisfied, we also obtain the BRST
transformation $s_b$ for the ghost field because the expansion for
${\cal F}^{(c)}$ of (9) becomes: ${\cal F}^{(c)} (x, \bar\theta) =
C + \bar \theta (s_b C)$.

We concentrate now on the explicit computation of the coefficients
of the 2-form differential $(dx^\mu \wedge d\bar\theta)$. The
final form of this expression is
\begin{equation}
i (dx^\mu \wedge d \bar\theta)\; \bar \Psi^{(c)}\; \Bigl (
\partial_{\bar\theta} {\cal B}_\mu^{(c)} - \partial_\mu {\cal
F}^{(c)} - i [ {\cal B}_\mu^{(c)}, {\cal F}^{(c)} ] \Bigr )\;
\Psi^{(c)}. \label{17}
\end{equation}
The restriction in (10) enforces the above coefficient to be zero.
This requirement leads to (with $\Psi^{(c)} (x, \bar\theta) \neq
0, \bar \Psi^{(c)} (x, \bar\theta) \neq 0$):
\begin{equation}
(R_\mu - D_\mu C) - i\; \bar\theta \;[ D_\mu B_1 + i (R_\mu \times
C) ] = 0, \label{18}
\end{equation}
which implies that $R_\mu = D_\mu C$. Setting equal to zero the
$\bar\theta$ part of the above equation, entails upon the
restriction $D_\mu [B_1 + \frac{1}{2} (C \times C)] = 0$ which is
readily satisfied due to the value of $B_1$ quoted in (16).

The most important piece of our present computation  is the
computation of the coefficient of the 2-form differential $(dx^\mu
\wedge dx^\nu)$ from the l.h.s. As is evident from (13), with a
little bit of algebra, the first term becomes:
\begin{equation}
\frac{i}{2} (dx^\mu \wedge dx^\nu)\; \bar\Psi^{(c)}\;
 \bigl ( \partial_\mu {\cal B}_\nu^{(c)} - \partial_\nu {\cal
 B}_\mu^{(c)} + i\; [ {\cal B}_\mu^{(c)}, {\cal B}_\nu^{(c)} ] \bigr
 )\; \Psi^{(c)}. \label{19}
\end{equation}
Substituting the explicit expressions for the expansions in (9),
we obtain the following form of the above equation
\begin{equation}
\frac{i}{2} (dx^\mu \wedge dx^\nu) \; \Bigl (\; \bar\psi (x)
F_{\mu\nu}  \psi (x) + i\; \bar\theta\; \bigl [ A_{\mu\nu} + i
B_{\mu\nu} \bigr ]\; \Bigr ), \label{20}
\end{equation}
where the explicit forms of $A_{\mu\nu}$ and $B_{\mu\nu}$, are
\begin{equation}
A_{\mu\nu} = \bar\psi (x)\; \bigl (\partial_\mu R_\nu -
\partial_\nu R_\mu + i [A_\mu, R_\nu] - i [A_\nu, R_\mu] \bigr )\;
\psi (x), \label{21}
\end{equation}
\begin{equation}
B_{\mu\nu} = \bar\psi (x)\; F_{\mu\nu}\; b_1 \cdot T  + b_2 \cdot
T \; F_{\mu\nu}\; \psi (x). \label{22}
\end{equation}
It is straightforward to note that the first term of (20) matches
with the r.h.s. of the restriction in (10). With the substitution
of $R_\mu = D_\mu C$, we obtain $A_{\mu\nu} = i (F_{\mu\nu} \times
C)$. Ultimately, setting the $\bar\theta$ part of (20) equal to
zero, leads to the following relationship \footnote{It will be
noted that the horizontality condition: $\tilde F_{|(c)}^{(2)} =
F^{(2)}$, where $\tilde F_{|(c)}^{(2)} = \tilde d \tilde
A^{(1)}_{|c} + i \tilde A^{(1)}_{|c}  \wedge \tilde A^{(1)}_{|c}$
and $F^{(2)} = d A^{(1)} + i A^{(1)} \wedge A^{(1)}$, leads to the
computation of the l.h.s. as: $(1/2) (dx^\mu \wedge dx^\nu) [
F_{\mu\nu} + i \bar \theta (F_{\mu\nu} \times C) ]$ whereas the
r.h.s. is $(1/2) (dx^\mu \wedge dx^\nu) (F_{\mu\nu})$ {\it alone}.
Here one does not set, the coefficient of $\bar\theta$ part of the
above equation, equal to zero because that would lead to an absurd
result: $(F_{\mu\nu} \times C) = 0$ (which is {\it not} the case
for our present 4D 1-form interacting non-Abelian gauge theory).
One circumvents this problem by stating that the kinetic energy
term $- (1/4)\; F^{\mu\nu} \cdot F_{\mu\nu}$ of the Lagrangian
density remains invariant (see, e.g., [3,4]) if $F_{\mu\nu} \to
F_{\mu\nu} + i \bar \theta (F_{\mu\nu} \times C)$. It should be
emphasized that such kind of problem does not arise in our present
attempt to derive the nilpotent (anti-)BRST symmetry
transformations with the gauge invariant restriction (10).}
\begin{equation}
\bar \psi (x)\;  (F_{\mu\nu} \times C) \;\psi (x) + \bar \psi
(x)\; F_{\mu\nu}\; b_1 \cdot T + b_2 \cdot T \;F_{\mu\nu}\; \psi
(x) = 0. \label{23}
\end{equation}
The above equation can be seen to be readily satisfied if we
choose $ b_1 \cdot T = - (C \cdot T) \psi (x)$ and $b_2 \cdot T =
- \bar \psi (x) (C \cdot T)$. With the help of these values, it
can be seen that the expansion for the matter superfields in (9)
become
\begin{eqnarray}
&&\Psi^{(c)} (x, \bar \theta) = \psi (x)\; + \;\bar \theta
\;(\tilde s_b \psi (x)), \nonumber\\ && \bar \Psi^{(c)} (x,
\bar\theta) = \bar \psi (x)\; + \;\bar\theta \; (\tilde s_b \bar
\psi (x)).\label{24}
\end{eqnarray}
The above equation provides the geometrical interpretation for the
on-shell nilpotent BRST transformation $\tilde s_b$ (and the
corresponding on-shell nilpotent BRST charge $\tilde Q_b$) as the
translational generator $(\partial/\partial \bar\theta)$ along the
Grassmannian direction $\bar\theta$ of the (4, 1)-dimensional
chiral supermanifold (cf. equation (8)). In fact, the process of
translation of the chiral matter superfields $\Psi^{(c)} (x,
\bar\theta) $ and $\bar\Psi^{(c)} (x, \bar\theta)$ along the
Grassmannian direction $\bar\theta$ results in the internal BRST
transformation $\tilde s_b$ on the corresponding local matter
fields $\psi (x)$ and $\bar\psi (x)$ of the Lagrangian density (4)
for the ordinary 4D theory.

The above interpretation of the BRST transformation $\tilde s_b$
(and the corresponding generator $\tilde Q_b$) is valid for all
the other superfields of (9). In this connection, it will be noted
that we have already computed $B_1 = - (1/2) (C \times C)$ and
$R_\mu = D_\mu C$ from the restriction (10). However, we have {\it
not} been able to say anything about the secondary field $B_2$,
present in the expansion of $\bar {\cal F}^{(c)}$. At this
juncture, the equation of motion $B = - (\partial_\mu A^\mu)$,
derived from the Lagrangian density (1), comes to our help as we
have the freedom to choose $B_2 \equiv B = - (\partial_\mu
A^\mu)$. All he above values, finally, imply the following
expansions for the chiral superfields defined in (9), namely;
\begin{eqnarray}
{\cal B}^{(c)} (x, \bar\theta) &=& A_\mu (x)\; + \;\bar\theta\;
(\tilde s_b A_\mu (x)), \nonumber\\ {\cal F}^{(c)} (x, \bar\theta)
&=& C (x) \; + \; \bar\theta\; (\tilde s_b C (x)), \nonumber\\
\bar {\cal F}^{(c)} (x, \bar\theta) & = & \bar C (x) \; + \;
\bar\theta\; (\tilde s_b \bar C (x)), \label{25}
\end{eqnarray}
which retain the geometrical interpretation of $\tilde s_b$ (as
well as $\tilde Q_b$) as the translational generator along the
Grassmannian direction $\bar\theta$ of the chiral supermanifold.
It will be noted that this conclusion was also drawn after (24).
In other words, the local internal BRST transformations $\tilde
s_b$ for the local basic fields $(A_\mu (x), C (x), \bar C (x))$
of the Lagrangian density (1) is equivalent to the translation of
the corresponding chiral superfields $({\cal B}_\mu^{(c)} (x,
\bar\theta)$, ${\cal F}^{(c)} (x, \bar\theta)$, $\bar {\cal
F}^{(c)} (x, \bar\theta)$ along the Grassmannian direction
$\bar\theta$ of the (4, 1)-dimensional chiral super sub-manifold
of the general (4, 2)-dimensional supermanifold.

\section{Off-shell nilpotent (anti-)BRST symmetry transformations:
superfield formalism}

In this section, we shall derive the off-shell nilpotent symmetry
transformations for {\it all} the fields of the (anti-)BRST
invariant Lagrangian densities (6) and (7) by invoking the same
restriction on the matter superfields as quoted in (10) but
defined on the general (4, 2)-dimensional supermanifold:
\begin{equation}
\bar \Psi (x, \theta, \bar\theta) \;\tilde {\cal D}\; \tilde {\cal
D}\; \Psi (x, \theta, \bar\theta)\; = \;\bar \psi (x) \; D\; D\;
\psi (x), \label{26}
\end{equation}
where all the superfields and super covariant derivatives  are
parameterized by four $x^\mu$ (with $\mu = 0, 1, 2, 3)$) spacetime
coordinates and a pair of Grassmannian variables $\theta$ and
$\bar\theta$. For instance, in the definition of the super
covariant derivative $\tilde {\cal D} = \tilde d + i \tilde
A^{(1)}$, the individual terms are as follows
\begin{eqnarray}
 \tilde d &=&\; dx^\mu\; \partial_\mu \;+ \;d \theta
\;\partial_\theta \;+ \;d \bar\theta \;\partial_{\bar\theta},
\nonumber\\ \tilde A^{(1)} &=& dx^\mu \;{\cal B}_\mu (x, \theta,
\bar\theta) + d \theta \;\bar {\cal F} (x, \theta, \bar\theta) + d
\bar\theta \; {\cal F} (x, \theta, \bar\theta). \label{27}
\end{eqnarray}
The super expansions for the multiplet fields ${\cal B}_\mu, {\cal
F}, \bar {\cal F}$, in terms of the basic fields $A_\mu, C, \bar
C$ as well as the secondary fields $R_\mu, \bar R_\mu, S_\mu, B_1,
\bar B_1, B_2, \bar B_2, s , \bar s$ , on the (4, 2)-dimensional
supermanifold, are [3,4]
\begin{eqnarray}
{\cal B}_\mu (x, \theta, \bar \theta) &=& A_\mu (x) + \theta\;
\bar R_\mu (x) + \bar \theta \;R_\mu (x) + i\; \theta\;
\bar\theta\; S_\mu (x), \nonumber\\ {\cal F} (x, \theta,
\bar\theta) &=& C (x) + i\; \theta\; \bar B_1 (x) + i \;\bar
\theta \;B_1 (x) + i\; \theta\; \bar\theta \;s (x), \nonumber\\
\bar {\cal F} (x, \theta, \bar\theta) & = & \bar C (x) + i
\;\theta \;\bar B_2 (x) + i \;\bar \theta\; B_2 (x) + i\; \theta\;
\bar\theta\; \bar s (x), \label{28}
\end{eqnarray}
where all the fields, in the above, are Lie algebra valued. In
other words, for the sake of brevity, we have taken the notations
$ {\cal B}_\mu = {\cal B}_\mu \cdot T, B_1 = B_1 \cdot T$, etc. In
the limit, $(\theta, \bar\theta) \to 0$, we retrieve all the basic
local gauge and (anti-)ghost fields of the Lagrangian densities
(1), (4),  (6) and (7). On the r.h.s. of the above expansion, we
can see that the fermionic and bosonic fields (and their degrees
of freedom) do match. The super expansions for the fermionic
matter superfields $( \Psi (x, \theta, \bar\theta, \bar \Psi (x,
\theta, \bar\theta)$ in (26), are as follows
\begin{eqnarray}
\Psi (x, \theta, \bar\theta) &=& \psi (x) + i \theta (\bar b_1
\cdot T) (x) + i \bar\theta (b_1 \cdot T) (x) + i \theta
\bar\theta (f \cdot T) (x), \nonumber\\ \bar \psi (x, \theta,
\bar\theta) & = & \bar \psi (x) + i \theta (\bar b_2 \cdot T) (x)
+ i \bar\theta (b_2 \cdot T) (x) + i \theta \bar\theta (\bar f
\cdot T) (x), \label{29}
\end{eqnarray}
where, it would be noted that, all the secondary fields are Lie
algebra valued but the Dirac fields (and corresponding
superfields) are not Lie algebra valued as is the case for these
fields in the Lagrangian densities.

It is clear that the r.h.s. of (26) (as discussed earlier), is
equal to an $SU(N)$  gauge invariant quantity $i \bar\psi (x)
F^{(2)} \psi (x)$ where the ordinary 2-form $F^{(2)} =
\frac{1}{2!} (dx^\mu \wedge dx^\nu)(\partial_\mu A_\nu -
\partial_\nu A_\mu + i A_\mu \times A_\nu)$. The latter contains
only a single wedge product of 2-form differentials (i.e. $(dx^\mu
\wedge dx^\nu)$) constituted by the spacetime variables alone.
However, the l.h.s. would produce all possible differentials of
2-form defined on the (4, 2)-dimensional supermanifold. To check
this statement, let us first expand the l.h.s. of the gauge
invariant restriction (26), in an explicit manner, as
\begin{eqnarray}
&& + (dx^\mu \wedge dx^\nu)\; \bar \Psi \;\bigl [ (\partial_\mu +
i {\cal B}_\mu)\;(\partial_\nu + i {\cal B}_\nu) \bigr ]\; \Psi
\nonumber\\&& - (d\theta \wedge d\theta)\; \bar \Psi\; \bigl [
(\partial_\theta + i \bar {\cal F}) (\partial_\theta + i \bar
{\cal F} \; \bigr ]\;\Psi  \nonumber\\ && - (d \bar\theta \wedge d
\bar\theta)\; \bar \Psi \;\bigl [ (\partial_{\bar\theta} + i {\cal
F}) (\partial_{\bar\theta} + i {\cal F}) \bigr ]\; \Psi\nonumber\\
&& - (d\theta \wedge d\bar\theta)\; \bar \Psi\; \bigl [
(\partial_{\bar\theta} + i {\cal F}) \; (\partial_\theta + i \bar
{\cal F}) + (\partial_\theta + i \bar {\cal F})
(\partial_{\bar\theta} + i {\cal F})\bigr ]\;\Psi \nonumber\\ && -
(dx^\mu \wedge d\theta)\; \bar \Psi \; \bigl [ (\partial_\mu + i
{\cal B}_\mu) (\partial_\theta + i \bar {\cal F}) -
(\partial_\theta + i \bar {\cal F}) (\partial_\mu + i {\cal
B}_\mu) \bigr ]\;\Psi \nonumber\\ && - (dx^\mu \wedge d\bar\theta)
\; \bar \Psi \bigl [ (\partial_\mu + i {\cal B}_\mu)
(\partial_{\bar\theta} + i {\cal F}) - (\partial_{\bar\theta} + i
{\cal F}) (\partial_\mu + i {\cal B}_\mu) \bigr ]\; \Psi,
\label{30}
\end{eqnarray}
where the anticommutativity property of the matter superfield
$\bar \Psi$ with the Grassmannian variables $\theta$ and
$\bar\theta$ has been taken into account.

For algebraic convenience, it is useful to first compare the
coefficients of the differentials $(d\theta \wedge d\theta)$, $(d
\bar\theta \wedge d\bar\theta)$ and $(d \theta \wedge
d\bar\theta)$ from the l.h.s. and r.h.s. of the gauge invariant
restriction (26). It is obvious that, on the r.h.s., there are no
such differentials. Thus, we have to set the above coefficients
from the l.h.s. equal to zero. These requirements lead to the
following relationships
\begin{eqnarray}
&& \partial_\theta \bar {\cal F} + i \bar {\cal F} \bar {\cal F} =
0 \Rightarrow \partial_\theta \bar {\cal F} + \frac{i}{2} \{ \bar
{\cal F}, \bar {\cal F} \} = 0, \nonumber\\ &&
\partial_{\bar\theta} {\cal F} + i {\cal F} {\cal F} = 0
\Rightarrow \partial_{\bar\theta} {\cal F} + \frac{i}{2} \{ {\cal
F}, {\cal F} \} = 0, \nonumber\\ && \partial_\theta {\cal F} \;+\;
\partial_{\bar\theta} \bar {\cal F}\; + \;i \;\{ {\cal F}, \bar {\cal F} \}
= 0, \label{31}
\end{eqnarray}
when $\Psi (x,\theta,\bar\theta) \neq 0,$ $ \bar \Psi (x,\theta,
\bar\theta) \neq 0$. The above conditions lead to the following
expressions for the secondary fields in terms of the basic fields:
\begin{eqnarray}
&& \bar B_2 = - \frac{1}{2} (\bar C \times \bar C),\; \qquad \bar
s = - i (B_2 \times \bar C), \nonumber\\ && \bar B_2 \times \bar C
= 0,\; \quad \bar C \times \bar s = i (B_2 \times \bar B_2),
\nonumber\\ && B_1 = - \frac{1}{2} (C \times C),\; \qquad s = i
(\bar B_1 \times C), \nonumber\\&& B_1 \times C = 0,\; \qquad (C
\times s) = i (B_1 \times \bar B_1), \nonumber\\ && \bar B_1 + B_2
= - (C \times \bar C), \nonumber\\&& C \times \bar s + s \times
\bar C = i (B_1 \times \bar B_2 - \bar B_1 \times B_2),
\nonumber\\ && s = i (C \times B_2 - B_1 \times \bar C),\; \qquad
\bar s = i ( C \times \bar B_2 - \bar B_1 \times \bar C).
\label{32}
\end{eqnarray}
This equation (i.e. (32)) shows that explicit values of $B_1, s,
\bar s$ and $\bar B_2$, in terms of the (anti-)ghost fields and
auxiliary fields, can be computed and the rest of the above
relations are consistent. To see the latter statement clearly, we
have to set equal to zero the coefficients of the differentials
$(dx^\mu \wedge d\theta)$ and $(dx^\mu \wedge d\bar\theta)$. These
conditions, for $\Psi \neq 0$ and $\bar \Psi \neq 0$, lead to
\begin{eqnarray}
\partial_\mu \bar {\cal F} - \partial_\theta {\cal B}_\mu +
i [{\cal B}_\mu,  \bar {\cal F}] &=& 0, \nonumber\\
\partial_\mu {\cal F} - \partial_{\bar\theta} {\cal B}_\mu  + i
[{\cal B}_\mu , {\cal F}] &=& 0. \label{33}
\end{eqnarray}
The outcome of the above conditions is listed below
\begin{eqnarray}
&& R_\mu = D_\mu C, \qquad \bar R_\mu = D_\mu \bar C, \qquad D_\mu
\bar B_2 + \bar R_\mu \times \bar C = 0, \nonumber\\ && S_\mu =
D_\mu B_2 + R_\mu \times \bar C \equiv D_\mu \bar B_1 + \bar R_\mu
\times C, \qquad D_\mu B_1 + R_\mu \times C = 0, \nonumber\\&&
D_\mu s = i (\bar R_\mu \times B_1 - R_\mu \times \bar B_1 + S_\mu
\times C), \nonumber\\ && D_\mu \bar s = i (\bar R_\mu \times B_2
- R_\mu \times \bar B_2 + S_\mu \times \bar C). \label{34}
\end{eqnarray}
It can be seen explicitly that all the above relationships are
consistent with one-another. It is very interesting to pinpoint
the fact that the restriction on the auxiliary fields of the
Lagrangian densities (6) and (7), advocated by Curci and Farrari
(i.e. $ B + \bar B = - (C \times \bar C)$ ) [21], automatically
emerges in our superfield approach if we identify $\bar B_1 = \bar
B$ and $B_2 = B$ (cf. (32)).

Finally, we concentrate on the computation of the coefficient of
$(dx^\mu \wedge dx^\nu)$ from the l.h.s. of the gauge-invariant
restriction (26). This can be explicitly expressed, after some
algebraic simplification, as
\begin{equation}
\frac{i}{2} (dx^\mu \wedge dx^\nu)\; \bar \Psi (x, \theta,
\bar\theta)\; \bigl (\partial_\mu {\cal B}_\nu - \partial_\nu
{\cal B}_\mu + i [{\cal B}_\mu, {\cal B}_\nu ] \bigr )\; \Psi (x,
\theta, \bar\theta). \label{35}
\end{equation}
We have to use, in the above, the super expansion of ${\cal
B}_\mu, \Psi, \bar \Psi$ from equations (28) and (29). Finally, we
obtain the following expression\footnote{It should be noted that,
in the horizontality condition $\tilde F^{(2)} = F^{(2)}$, the
analogue of (35) from the l.h.s. yields $\frac{i}{2} (dx^\mu
\wedge dx^\nu) [F_{\mu\nu} + i \theta (F_{\mu\nu} \times \bar C) +
i \bar\theta (F_{\mu\nu} \times C) - \theta \bar\theta (F_{\mu\nu}
\times B + F_{\mu\nu} \times C \times \bar C)]$. But, the r.h.s.
is only $\frac{1}{2} (dx^\mu \wedge dx^\nu) F_{\mu\nu}$. One does
not set here the $\theta, \bar\theta$ and $\theta\bar\theta$ parts
equal to zero because these lead to absurd results. Rather, one
gets rid of this problem by stating that the kinetic energy term
$- \frac{1}{4} F_{\mu\nu} \cdot F^{\mu\nu}$ remains invariant
under $F_{\mu\nu} \to F_{\mu\nu} + i \theta (F_{\mu\nu} \times
\bar C) + i \bar\theta (F_{\mu\nu} \times C) - \theta \bar\theta
(F_{\mu\nu} \times B + F_{\mu\nu} \times C \times \bar C)$ (see,
e.g., [3,4]).}
\begin{equation}
\frac{i}{2} (dx^\mu \wedge dx^\nu)\; \Bigl [ \bar\psi F_{\mu\nu}
\psi + i\; \theta \;L_{\mu\nu} + i\; \bar\theta\; M_{\mu\nu} + i\;
\theta\;\bar\theta\;N_{\mu\nu} \Bigr ], \label{36}
\end{equation}
where the expressions for $L_{\mu\nu}, M_{\mu\nu}$ and
$N_{\mu\nu}$, in full blaze of glory, are
\begin{eqnarray}
L_{\mu\nu} &=& \bar b_2 F_{\mu\nu} \psi - \bar\psi F_{\mu\nu} \bar
b_1 - \bar \psi (F_{\mu\nu} \times \bar C) \psi, \nonumber\\
M_{\mu\nu} &=& b_2 F_{\mu\nu} \psi - \bar \psi F_{\mu\nu} b_1 -
\bar\psi (F_{\mu\nu} \times C) \psi, \nonumber\\ N_{\mu\nu} &=&
\bar f F_{\mu\nu}\psi + \bar\psi F_{\mu\nu} f - i \bar\psi
(F_{\mu\nu} \times \bar C) b_1 + i \bar \psi (F_{\mu\nu} \times C)
\bar b_1 \nonumber\\ &+& i \bar \psi [F_{\mu\nu} \times (B_2 + C
\times \bar C)] \psi + i \bar b_2 F_{\mu\nu} b_1 \nonumber\\ &+& i
\bar b_2 (F_{\mu\nu} \times C) \psi - i b_2 F_{\mu\nu} \bar b_1 -
i b_2 (F_{\mu\nu} \times \bar C) \psi. \label{37}
\end{eqnarray}
It is straightforward to check that the first term of (36) does
match with the explicit computation of the r.h.s. (i.e. $ i \bar
\psi F^{(2)} \psi$)  of the gauge invariant restriction (26). This
implies immediately that $L_{\mu\nu}, M_{\mu\nu}$ and $N_{\mu\nu}$
must be set equal to zero. It is not very difficult to check that
$L_{\mu\nu} = 0$ and $M_{\mu\nu} = 0$ demand the following
expressions for $b_1, b_2, \bar b_1, \bar b_2$; namely,
\begin{equation}
\bar b_2 = - \bar \psi (\bar C \cdot T),\; \quad \bar b_1 = -
(\bar C\cdot T) \psi, \;\quad b_2 = - \bar\psi (C \cdot T)\; \quad
b_1 = - (C\cdot T) \psi. \label{38}
\end{equation}
A few points, regarding the above solutions, are in order. First,
a close look at the equation $L_{\mu\nu} = 0$ shows that $\bar
b_2$ and $ \bar b_1$ should be proportional to $\bar\psi$ and
$\psi$, respectively. Second, to maintain the bosonic nature of
$\bar b_2$ and $\bar b_1$, it is essential that a single fermion
should be brought in, together with $\bar \psi$ and $\psi$.
Finally, the Lie algebra valuedness of $\bar b_2$ and $\bar b_1$
requires that $(\bar C \cdot T)$ should be brought in for the
precise cancellation so that we obtain $L_{\mu\nu} = 0$. Precisely
similar kinds of arguments go into the determination of the
solutions to the equation $M_{\mu\nu} = 0$.

Finally, we would like to devote time on finding out the solutions
to the condition $N_{\mu\nu} = 0$. First of all, it can be seen
that we can exploit the values from (38) to simplify $N_{\mu\nu}$.
For instance, it can be noted that
\begin{equation}
- i \bar\psi (F_{\mu\nu} \times \bar C) b_1 - i b_2 (F_{\mu\nu}
\times \bar C) \psi = i \bar\psi \{ F_{\mu\nu} \times \bar C, C \}
\psi \equiv i \bar\psi (F_{\mu\nu} \times C \times \bar C) \psi,
\label{39}
\end{equation}
and exactly similar kind of counter terms (present in
$N_{\mu\nu}$)
\begin{equation}
i \bar \psi\; (F_{\mu\nu} \times C)\; \bar b_1 + i \;\bar b_2\;
(F_{\mu\nu} \times C) \;\psi \equiv - i\; \bar \psi\;
(F_{\mu\nu}\times C \times \bar C) \;\psi, \label {40}
\end{equation}
add to zero. Out of the remaining terms, it can be seen that
\begin{equation}
i \bar b_2 F_{\mu\nu} b_1 - i b_2 F_{\mu\nu} \bar b_1 = -
\frac{i}{2} \bar\psi (F_{\mu\nu} \times C \times \bar C) \psi.
\label{41}
\end{equation}
Thus, ultimately, we obtain the following surviving terms in
$N_{\mu\nu}$
\begin{equation}
\bar f\; F_{\mu\nu}\; \psi + \bar \psi\; F_{\mu\nu}\; f + i
\;\bar\psi \;\Bigl (F_{\mu\nu} \times \bigl (B_2 + \frac{1}{2} C
\times \bar C \bigr ) \Bigr )\; \psi, \label{42}
\end{equation}
which immediately allows us to choose (with identification $B_2 =
B$)
\begin{equation}
f = -\;i \;(B + \frac{1}{2} C \times \bar C)\; \psi, \qquad \bar f
= i\; \bar \psi \;(B + \frac{1}{2} C \times \bar C), \label{43}
\end{equation}
so that $N_{\mu\nu} = 0$. Finally, the super expansions in (28)
and (29), after insertion of the values from (32), (34), (38) and
(43), become
\begin{eqnarray}
{\cal B}_\mu (x, \theta, \bar\theta) &=& A_\mu (x) + \theta
(s_{ab} A_\mu (x)) + \bar\theta (s_b A_\mu (x)) + \theta
\bar\theta (s_b s_{ab} A_\mu (x)), \nonumber\\ {\cal F} (x,
\theta, \bar\theta) &=& C (x) + \theta\; (s_{ab} C  (x)) +
\bar\theta\; (s_b C(x)) + \theta \;\bar\theta \;(s_b s_{ab} C
(x)), \nonumber\\ \bar {\cal F} (x, \theta, \bar\theta) &=& \bar C
(x) + \theta \;(s_{ab} \bar C(x)) + \bar\theta \;(s_b \bar C (x))
+ \theta\; \bar\theta \;(s_b s_{ab} \bar C (x)), \nonumber\\ \Psi
(x, \theta, \bar\theta) & = & \psi (x) + \theta \;(s_{ab} \psi
(x)) + \bar\theta \;(s_b \psi (x)) + \theta \;\bar\theta \;(s_b
s_{ab} \psi (x)), \nonumber\\ \bar \Psi (x, \theta, \bar\theta) &
= & \bar\psi (x) + \theta \;(s_{ab} \bar \psi (x)) + \bar\theta\;
(s_b \bar\psi (x)) + \theta \;\bar\theta \;(s_b s_{ab} \bar\psi
(x)). \label {44}
\end{eqnarray}
The above expansions, once again, demonstrate the geometrical
interpretation of the (anti-)BRST symmetry transformations (and
their corresponding generators $Q_{(a)b}$) as the translational
generators along the Grassmannian directions $(\theta)\bar\theta$
of the (4, 2)-dimensional general supermanifold. Mathematically,
the nilpotency property ($s_{(a)b}^2 = 0, Q_{(a)b}^2 = 0$), the
anticommutativity property ($s_b s_{ab} + s_{ab} s_b = 0, Q_b,
Q_{ab} + Q_{ab} Q_b = 0)$, etc., can be expressed in terms of the
translational generators as
\begin{eqnarray}
&&s_b\;\; \Leftrightarrow \;\;Q_b \;\;\Leftrightarrow\;\;
\mbox{Lim}_{\theta \to 0} \frac{\partial}{\partial \bar\theta},
\quad s_{ab}\;\; \Leftrightarrow\;\; Q_{ab} \;\;\Leftrightarrow
\;\; \mbox{Lim}_{\bar\theta \to 0} \frac{\partial}{\partial
\theta}, \nonumber\\ && s_{b}^2 = 0\;\;\; \Leftrightarrow\;\;\;
Q_{b}^2 = 0\;\;\; \Leftrightarrow \;\;\;\Bigl ( \mbox{Lim}_{\theta
\to 0}\; \frac{\partial}{\partial\bar\theta} \Bigr )^2 = 0,
\nonumber\\&& s_{ab}^2 = 0 \;\;\;\Leftrightarrow\;\;\; Q_{ab}^2 =
0 \;\;\;\Leftrightarrow \;\;\;\Bigl ( \mbox{Lim}_{\bar\theta \to
0}\; \frac{\partial}{\partial \theta} \Bigr )^2 = 0, \nonumber\\
&& s_b s_{ab} + s_{ab} s_b = 0 \;\;\;\Leftrightarrow \;\;\;Q_b
Q_{ab} + Q_{ab} Q_b = 0 \;\;\;\Leftrightarrow \nonumber\\ && \Bigl
( \mbox{Lim}_{\bar\theta \to 0} \frac{\partial} {\partial\theta}
\Bigr ) \; \Bigl ( \mbox{Lim}_{\theta \to 0}
\frac{\partial}{\partial \bar\theta} \Bigr ) + \Bigl (
\mbox{Lim}_{\theta \to 0} \frac{\partial}{\partial \bar\theta}
\Bigr )\; \Bigl ( \mbox{Lim}_{\bar\theta \to 0}
\frac{\partial}{\partial \theta} \Bigr ) = 0. \label{45}
\end{eqnarray}
This establishes the geometrical interpretations for all the
mathematical properties associated with $s_{(a)b}$ and $Q_{(a)b}$.

\section{Conclusions}

One of the central results of our present investigation is the
precise derivation of the full set of on-shell as well as
off-shell nilpotent (anti-)BRST symmetry transformations
associated with all the fields of a given 1-form 4D interacting
non-Abelian gauge theory in the superfield formulation. These
symmetries emerge from a {\it single} gauge (i.e. BRST) invariant
restriction (cf. (10) and (26)) on the matter superfields defined
on the appropriate supermanifolds. The above restriction is a bold
statement that the physical (i.e. BRST invariant) quantities
should remain unaltered even in the presence of supersymmetric
(Grassmannian) coordinates that appear in the superfield approach
to BRST symmetries . This amounts to the requirement that all the
wedge products (and otherwise too) of the Grassmannian variables,
present in the definition of the above BRST invariant quantities
(cf. (10), (26)), should be set equal to zero because the r.h.s.
of the above quantities are without them.

The above cited gauge (i.e. BRST) invariant quantities originate
from the key properties associated with the (super) covariant
derivatives and their intimate connections with the definition of
the curvature forms on the supermanifolds. Some of the striking
similarities and key differences between the horizontality
condition and our gauge invariant condition are as follows. First,
both of them primarily owe their origin to the (super)
cohomological operators $\tilde d $ and $d$. Second, the
geometrical origin and interpretations for the (anti-)BRST charges
(and the nilpotent symmetry transformations they generate) remain
intact for the validity of both the conditions on the superfields.
Third, whereas the horizontality condition is an $SU(N)$ covariant
restriction (because $F^{(2)} \to U F^{(2)} U^{-1}$ where $U \in
SU(N)$), the other condition, as the name suggests, is an $SU(N)$
gauge invariant condition. Fourth, the gauge (i.e. BRST) invariant
restrictions in (10) and (26) are basically the generalization of
the horizontality condition {\it itself} where the matter fields
(and the corresponding superfields) have been brought into the
picture so that these combinations could become the gauge (i.e.
BRST) invariant quantities. Finally, there is a very crucial
logical (as well as mathematical) difference between the
horizontality restriction and the gauge invariant restrictions in
(10) and (26). This has been elaborated clearly and cogently in
the footnotes before equations (23) and (36) of our present paper.

It is worthwhile to mention that the gauge invariant restrictions
in (10) and (26) are superior to (i) the horizontality condition
applied in the context of the usual superfield formulation [1-7],
and (ii) the consistent extensions of the horizontality condition
in the case of the augmented superfield formalism [8-16]. This is
due to the fact that (i) whereas the horizontality condition
(modulo some logical mathematical questions) leads to the
derivation of the nilpotent symmetry transformations for the gauge
and (anti-)ghost fields, our gauge invariant restrictions yield
{\it all} the symmetry transformations for {\it all} the fields,
and (ii) whereas in the augmented superfield approach, the
horizontality condition and the additional restriction(s) are
exploited separately and independently, one obtains all the
nilpotent (anti-)BRST symmetry transformations for all the fields
in one stroke from the gauge invariant restrictions (exploited in
(10) and (26) for the appropriately chosen matter superfields).

The highlights of our present endeavour could be enumerated as
follows. First of all, the restrictions in (10) and (26) are
physically as well as aesthetically more appealing because they
are BRST invariant. Second, these gauge (i.e. BRST) invariant
restrictions on the superfields are more economical because they
produce all the nilpotent symmetry transformations for all the
fields of a given 1-form interacting (non-)Abelian gauge theory in
one stroke. Finally, these restrictions on superfields have very
sound mathematical basis at the conceptual level as well as at the
algebraic level. Thus, in our entire computation, the thread of
logical coherence runs through everywhere.

It would be interesting to extend our prescription (e.g. equations
(10) and (26)) to a different set of interacting systems so that,
the idea proposed in our present investigation, can be put on a
firmer footing. For instance, one can check the validity of the
analogues of the restrictions (10) and (26) in the context of the
interacting $U(1)$ gauge theory where the charged complex scalar
fields couple to the $U(1)$ gauge field. It would be more
challenging to test the usefulness and sanctity of our idea in the
case of gravitational theories (see, e.g., [4] for earlier work)
where superfield formulation has been applied to derive the
nilpotent (anti-)BRST symmetries. These are some of the immediate
issues that are presently under investigation and our results
would be reported in our forthcoming future publications [22].

\end{document}